\documentclass[runningheads, a4paper]{llncs}

\usepackage{algorithm}
\usepackage{graphicx}
\usepackage{subcaption}
\usepackage{multicol}
\usepackage{xcolor}
\graphicspath{ {./images/} }
\usepackage[noend]{algpseudocode}
\usepackage{multirow}

\usepackage[utf8]{inputenc}

\usepackage{algorithm}
\usepackage{algpseudocode}
\usepackage{amsmath}
\usepackage[breaklinks=true]{hyperref}

\title{Porting numerical integration codes from CUDA to oneAPI: a case study\thanks{code available at https://github.com/marcpaterno/gpuintegration}}

\author{
		Ioannis Sakiotis\inst{1} \and
		Kamesh Arumugam\inst{2} \and
		Marc Paterno\inst{1,3}\and
		Desh Ranjan\inst{1}\and
		Bal{\v s}a Terzi\'c\inst{1}\and
		Mohammad Zubair\inst{1}
	}
	
	\authorrunning{Ioannis Sakiotis et al.}
	
	\institute{
		Old Dominion University, Norfolk, VA 23529, USA \and
		NVIDIA, Santa Clara, CA 95051-0952, USA \and
		Fermi National Accelerator Laboratory, Batavia, IL 60510
	}

\date{January 2023}

\begin{document}

\maketitle

\begin{abstract}
    
    We present our experience in porting optimized CUDA implementations to oneAPI. We focus on the use case of numerical integration, particularly the CUDA implementations of PAGANI and $m$-Cubes. We faced several challenges that caused performance degradation in the oneAPI ports. These  include differences in utilized registers per thread, compiler optimizations, and mappings of CUDA library calls to oneAPI equivalents. After addressing those challenges, we tested both the PAGANI and $m$-Cubes integrators on numerous integrands of various characteristics. To evaluate the quality of the ports, we collected performance metrics of the CUDA and oneAPI implementations on the Nvidia V100 GPU. We found that the oneAPI ports often achieve comparable performance to the CUDA versions, and that they are at most $10\%$ slower.     
    
\end{abstract}

\section{Introduction}

Historically, general-purpose GPU programming has been characterized by divergent architectures and programming models.
A lack of widely adopted common standards led to the development of different ecosystems comprised of compilers and tools that were practically exclusive to specific GPU architectures.
Most importantly, the emergent architectures themselves were not compatible with all ecosystems.
Portability could only be achieved through the maintenance of multiple code bases.
Traditionally, the proprietary CUDA programming model has been the most popular but is exclusively targeted to Nvidia GPUs.

In the absence of universally adopted standards, a viable solution for achieving general portability is to rely on platform-agnostic programming models that target multiple architectures via a unifying interface.
This enables the execution of a single code base across various architectures.
These programming models would ideally enable the utilization of platform-specific low-level features on their native hardware.
This would allow highly-optimized implementations in such portable programming models to remain competitive with platform-specific alternatives.
Without these capabilities, use cases with extreme performance requirements would disqualify the use of such portable models.

The need for performant multi-platform execution is only increasing with the emergence of exascale supercomputers such as Frontier and Aurora that do not carry Nvidia GPUs.
Projects requiring computing cores at that scale must develop new software solutions compatible with non-Nvidia GPUs or port existing CUDA implementations without significant loss of performance.

Portable programming models such as RAJA, Kokkos, and oneAPI have been in development and are already available for use.
These portable alternatives lack maturity when compared to proprietary alternatives.
As such, applications requiring portable solutions must be evaluated to quantify any necessary concessions.

In this paper, we discuss the porting process of two numerical integration implementations, PAGANI and m-Cubes, from CUDA to oneAPI's Data Parallel C++ language.
The oneAPI ecosystem provides a suite of compilers, libraries, and software tools,  including the Data Parallel Conversion tool (DPCT) that automates the majority of the porting process.
Reliance on the C++ and SYCL standards as well as the capability to quickly port large CUDA implementations, places oneAPI at the forefront of the portability initiative.

We faced challenges during the porting process due to the lack of support for certain libraries utilized by the CUDA implementation.
For example, the CUDA implementation of PAGANI uses the Nvidia Thrust library to perform common parallel operations on the host side, such as inner product and min-max.
Even though there is a multitude of library options in oneAPI, we encountered difficulties with the DPCT mapping of Nvidia Thrust library calls, which were not fully supported on all backends.

We also observed performance degradation for the ported oneAPI implementations.
We conducted numerous experiments with integrands of various characteristics to identify the issues.
Most of these issues pertained to optimization differences between the NVCC and Clang compilers, and time differences when executing mathematical functions.
After addressing these challenges, the oneAPI ports were at most $10\%$ slower than the optimized CUDA versions.
We observe that the cases with the highest performance penalties for the oneAPI ports, require significantly more registers than the CUDA originals. 
This decreases the occupancy in the oneAPI implementation and causes performance degradation. When the number of registers is similar to the CUDA version, we observe penalties lower than $5\%$.

The remainder of this paper is structured as follows.
First, we provide background information on oneAPI and other portability solutions in section $2$.
Then, we discuss the two numerical integration CUDA implementations in section $3$.
Section $4$ details the porting process and challenges we faced using DPCT and the oneAPI platform.
In section $5$, we present a performance comparison of the CUDA and oneAPI implementations of PAGANI and m-Cubes.
We finish in section $6$ with a discussion of our conclusions regarding the oneAPI platform's viability and ease of use.
We demonstrate that the oneAPI implementation does not induce significant performance penalties and that it is a viable platform for attaining performance on Nvidia GPUs.

\section{Background}
There are multiple programming models targeting different architectures.
Among the most prominent, are OpenCL \cite{opencl},  \cite{sycl}, OpenACC \cite{openacc}, OpenMP \cite{openmp}, RAJA, Alpaka \cite{alpaka}, and Kokkos \cite{kokkos}.
The Khronos group was the first to address portability by developing the OpenCL standard to target various architectures.
The same group later followed with the SYCL standard.
SYCL  is a higher-level language that retained OpenCL features but significantly improved ease of use with the utilization of C++ and the adoption of a single-source model.
There are multiple implementations of  SYCL such as DPC++, ComputeCpp, HipSYCL, and triSYCL \cite{sycl_overview}.
DPC++ is conformant to the latest SYCL and C++ standards and is integrated into the oneAPI ecosystem \cite{alcf}.

\subsection{Intel oneAPI and DPC++}

The oneAPI toolkit was released in 2020, providing a programming platform with portability across multiple architectures at the core of its mission.
The toolkit included various tools along with the DPC++ language which was based on the SYCL and C++ standards \cite{dpcpp}.
The reliance on these open standards that are intended to evolve over time is one of the most attractive features of DPC++.
Such evolution is facilitated by DPC++ extensions with various features that can be later introduced to the standards after periods of experimentation.
Such examples include the use of \emph{Unified Memory} and \emph{filtered Device selectors}, which were missing from SYCL 1.2.1 but were later included in the SYCL 2020 standard.
DPC++ achieves execution platform portability through its use of SYCL and various backends (implemented as shared libraries) that interface with particular instruction sets such as PTX for Nvidia GPUs and SPIR-V for Intel devices.
It is worth noting that there is no reliance on OpenCL, which is instead one of several available backends.
As such, DPC++ implementations can target various CPUs, GPUs, and FPGAs.
This is a similar approach to Kokkos, Alpaka, and RAJA.

\subsection{CUDA-backend for SYCL}

While CUDA is the native and most performant programming model for Nvidia GPUs, Nvidia provided support to the OpenCL API \cite{opencl_cuda}.
As a result, non-CUDA implementations could be executed on Nvidia GPUs.
The ComputeCpp implementation of SYCL by CodePlay, provided such functionality through OpenCL, but its performance was not comparable to native CUDA as not all functionality was exposed \cite{computecpp}.

As such, CodePlay developed the CUDA backend for DPC++, which is part of the LLVM compiler project.
CUDA support is not enabled by default and is at an experimental stage.
To enable the backend, we must build the LLVM compiler project for CUDA.
This can be achieved through easy-to-follow instructions that involve CUDA-specific flags, and the use of \emph{clang++} instead \emph{dpcpp} to compile source code.
As a result, DPC++ code can generate PTX code by using CUDA directly instead of relying on the OpenCL backend.
This approach not only enables the use of Nvidia libraries and profiling tools with DPC++ but also the capability to theoretically achieve the same performance as CUDA.

\subsection{Related Work}

The oneAPI programming model may not be as mature as CUDA but the literature already includes several examples of utilizing DPC++.
The authors of \cite{stencil_code} validated the correctness of a DPC++ tsunami simulator ported from CUDA.
A Boris Particle Pusher port from an openMP version was discussed in \cite{boris_p_pusher}, where a DPC++ implementation was $10\%$ slower than the optimized original.
In \cite{costanzo1}, CUDA and DPC++ implementations of a matrix multiplication kernel were compared on different matrix sizes; the execution time on an Nvidia GPU was slower with DPC++ code by $7\%$ on small problem sizes but as much as $37\%$ on larger ones.
On the contrary, \cite{costanzo2} and \cite{costanzo3} included experiments where a DPC++ biological sequence alignment code showed no significant performance penalty compared to CUDA, and even a case of $14\%$ speedup.
Spare matrix-vector multiplication kernels and \emph{Krylov} solvers in \cite{linalg} reached $90\%$ of a CUDA version's bandwidth.
There were also cases with non-favorable performance for DPC++ ports.
In \cite{bioinf_kernel} a bioinformatics-related kernel performed twice as fast in CUDA and HIP than in DPC++.
In \cite{CASTANO2022120} DPC++ versions generally reported comparable performance to CUDA but there were multiple cases where the penalty ranged from $25-190\%$.

There seems to be a deviation in the attainable performance.
This is reasonable due to the variety of applications and the relatively early stage of development for the oneAPI ecosystem.
We also expect that the level of optimization in CUDA implementations is an important factor.
In our experience, highly optimized codes typically yield performance penalties in the range ($5-10\%$).
There are multiple cases displaying approximately $10\%$ penalty compared to native programming models.
This indicates that DPC++ can achieve comparable performance to CUDA, though careful tuning and additional optimizations may be needed.

\section{Numerical Integration Use Case}

Numerical integration is necessary for many applications across various fields and especially physics.
Important examples include the simulation of beam dynamics and parameter estimation in cosmological models \cite{physics_app} \cite{des} \cite{cosmosis}.
Even ill-behaving integrands (oscillatory, sharply peaked, etc.) can be efficiently integrated with modest computational resources, as long the integration space is low dimensional (one or two variables).
On the contrary, solving medium to high-dimensional integrands is often infeasible on standard computing platforms.
In such cases, we must execute on highly parallel architectures to achieve performance at scale.
There are a few  GPU-compatible numerical integration algorithms \cite{pagani} \cite{mcubes} \cite{twophase} \cite{kanzaki} \cite{zmc}.
Unfortunately, exploration of execution-platform portability has been limited, with CUDA being the most common choice.
Since CUDA is a proprietary language, such optimized implementations cannot be executed on non-Nvidia GPUs.
To our knowledge, the only mentions of potential portability in numerical integration libraries are found in \cite{pagani} where a Kokkos implementation of the PAGANI integrator is briefly mentioned to be in development and in \cite{mcubes} which compares the CUDA implementation of $m$-Cubes with an experimental Kokkos version.

\subsection{PAGANI}

PAGANI is a deterministic quadrature-based algorithm designed for massively parallel architectures.
The algorithm computes an integral by evaluating the quadrature rules, which are a series of weighted summations of the form $\sum_{i=1}^{f_{eval}} w_i \cdot f(x_i)$.
The computation involves an integrand function $f$ which we invoke at the $d$-dimensional points $x_i$.
Each point $x_i$ has a corresponding weight $w_i$ and there are $f_{eval}$ such points in the summation.
PAGANI computes an initial integral and error estimate, and it progressively improves its accuracy until reaching a user-specified threshold.
The accuracy improvement is achieved by applying the quadrature rules in smaller regions of the integration space and accumulating those values to get the integral estimate.

PAGANI, is a memory-hungry algorithm due to its rapid sub-division of the integration space; high-accuracy executions typically generate millions of regions for ill-behaving integrands.
This is a design choice that is greatly suited for parallel systems such as GPUs.
A significant component of the algorithm involves the careful classification of the regions, with the goal of eliminating certain classes of regions from the limited available memory.
While this is a critical step for the progression of the algorithm, it is not the most computationally intensive.

\begin{algorithm}
	\caption{PAGANI}
	\begin{algorithmic}[1]
		\Procedure{\sc{pagani-kernel}}{$f$, $n$, $d$, $R[n]$, $I[n]$, $E[n]$, $K[n]$, $q$}
		\For{each region $i$ to $n$  parallel}
		\State local[5] $\leftarrow 0$ \Comment{thread-local storage for evaluations of the five quad. rules}
		\If{thread 0}
		\State shared [$f_{eval}$] $\leftarrow 0$  \Comment{store function evalutions in shared mem.}
		\State $S \leftarrow$ \textsc{Init-shared-mem}($R$) \Comment{collection of info. on region processed}
		\EndIf

		\For{each $f_{id} < f_{eval}$ parallel}
		\State x[$d$] $\leftarrow$ \textsc{compute-eval-point}($S$, $d$, $q$)
		\State shared[$t_{id}$] $\leftarrow f(x)$
		\For{$i\leftarrow 0$ to $4$} \Comment{per quad-rule}
		\State local[$i$] $\mathrel{+}= f(x) \cdot q.weights[t_{id}]$
		\EndFor
		\EndFor

		\For{$i\leftarrow 0$ to $4$ parallel} \Comment{per quad-rule}
		\State local[$i$] $\leftarrow$ \textsc{Reduce}($local[i]$)
		\EndFor

		\If{thread 0}
		\State $S.k \leftarrow$ \textsc{Compute-split-axis}($shared, q, d$)
		\State ${S.e} \leftarrow$ \textsc{find-max-err}($local, q$)
		\State $S.i \leftarrow$ local[$0$]
		\State $I, E, K \leftarrow S.i, S.e, S.k$ \Comment{final results}
		\EndIf
		\EndFor
		\EndProcedure
	\end{algorithmic}
	\label{alg:classify_alg}
\end{algorithm}

In fact, it is the \textsc{Evaluate} method (listed in Algorithm $2$  of \cite{pagani}) that consistently takes more than $90\%$ of total execution time.
Its function is to compute an integral/error estimate for each region and select one of the dimensional axes for splitting.
 As such, it can be viewed as the core of PAGANI, both from an algorithmic and performance standpoint.

We describe the \textsc{Evaluate} method in Listing 1, which we will refer to as \textsc{Pagani-kernel}.
The input consists of the integrand $f$, which will be repeatedly evaluated within the boundaries of each region, the number of regions $n$, and the dimensionality $d$.
Then, we supply the region coordinates $R$ in the form of two lists $R.c$ (low boundary of a region on each axis) and $R.l$ (length of a region on each axis).
Next, we use the lists $I$ and $E$ to store the integral and error estimate of each region, while the list $K$ contains an index representing the axis that should be split in the future to improve the estimates.
Finally, we supply a collection of read-only lists through the parameter $q$.
 Each list in $q$ contains constant values used to generate and permute over the evaluation points and weights associated with the five quadrature rules.

Each thread group processes a single region $i$ and uses all threads in a group to parallelize the integrand function evaluations.
On lines $3$ to $6$ each thread initializes its variables, while thread zero initializes variables shared in the group.
The $local$ list at line $3$ has a single entry for each of the five quadrature rules.
All threads compute their assigned function evaluations for each quadrature rule and accumulate the results in the $local$ list.
Each of the five entries in $local$, contains the partial results computed by the thread for the respective quadrature rule summations.
$shared$ is an array in shared memory.
It is used for the temporary storage of function evaluations that need to be retained and not just accumulated.

At line $6$, thread zero of each thread group will cache the sub-region assigned to its thread group into the $S$ shared memory location.
$S$ contains the region boundaries, global boundaries of integration space, volume, longest dimensional axis in the boundaries of the region, integral/error-estimate, and axis to split.
All of these values are associated with a single region.

Then the for-loop at line $7$ executes each function evaluation $f_{id}$ in parallel.
At line $8$, each thread computes the integrand $f$ at a $d$-dimensional point $x$.
This involves reading $d$ values from one of the lists in $q$ (generators list) and scaling those values to the region boundaries of each axis (bounds stored in $S$).
Then the threads evaluate the integrand at line $9$ and store the result at the appropriate index in the list $shared$.
Once that is done, the threads use the weights associated with each quadrature rule (stored in $q.w$) to multiply by the function evaluation at line $11$.

Once the threads exit the for-loop of line $7$, all function evaluations have been completed.
At line $13$, the algorithm accumulates the results from all threads in the group, which yield five integral estimates per region.
At the end, thread zero will store the reduced data as the final estimates in line $18$.
Thread zero also computes the split axis at line $15$.
This requires reading two values from a list in $q$, and the serial iteration of several function evaluations which are stored in $shared$.
Thread zero performs a series of simple operations and then stores the index of the axis to split at the shared location $S.k$.
The error estimate is computed at line $16$ and involves determining the maximum difference between the estimates reported by four of the five quadrature rules.
This requires reading from certain lists in $q$ and the produced result is the error estimate for the integral estimate of that region.

The CUDA implementation was optimized for the Nvidia \emph{V100} GPU.
The kernel is launched in groups of $64$ threads and the loop at line $7$ in iterated in a strided fashion.
This allows the threads to coalesce accesses to global memory (reads in $q$) needed for the function evaluations.
Additionally, the kernel relies on the ``ldg" intrinsic for reading the arrays within $q$, suggesting to the compiler their placement in the read-only cache.

It is worth noting that this is a compute-bound kernel with hundreds and often thousands (depending on the integrand) of floating-point operations per byte.
While the algorithm typically minimizes branch divergence, it cannot be avoided entirely due to thread zero performing operations on shared memory and storing the final results in global memory.
Due to fast-access shared memory utilization, the implementation uses synchronization points in the following cases: prior to the loop in line $7$, between the main loop strides, and before reducing the data at line $13$.
The reduction utilizes warp-level primitives to accumulate results within the warp, but needs shared memory to reduce the values among the warps.

\subsection{$m$-Cubes}

$m$-Cubes is a probabilistic Monte Carlo algorithm based on the VEGAS integrator \cite{PETERLEPAGE1978192}.
It operates by randomizing the sample generation across the integration space to solve integrands and relies on the standard deviation of the Monte Carlo estimate to produce error estimates for the computation.
Just like VEGAS, $m$-Cubes utilizes importance and stratified sampling to accelerate the Monte Carlo rate of convergence.
Importance sampling works by progressively approximating a step-wise function that mirrors the shape of the integrand.
The step-wise function, allows the algorithm to sample from a probability distribution that concentrates the function evaluations where the integrand varies the most.
This approach achieves the adaptive behavior that is necessary for numerical integrators to be efficient in high-dimensional spaces.

In terms of stratified sampling, $m$-Cubes partitions the integration space into $m$ sub-cubes that are sampled separately.
This improves the convergence rate and eliminates statistical errors and inefficient sampling that could occur on sharp-peaked integrands.
The sub-division resolution is dictated by a user-specified number of samples per iteration.
$m$-Cubes partitions the integration space such that a $2$-point Monte Carlo integration can be executed on each partition while generating approximately $n$ samples.

We describe the main kernel in Listing 2, though it is also detailed in \cite{mcubes}.
The implementation relies on many thread groups of size $x$, requiring a total $\frac{m}{x}$ such groups where $m$ is the number of sub-cubes resulting from the initial sub-division.
Each thread processes its own sub-cube batch of size $s$.

\begin{algorithm}
	\caption{$m$-Cubes}
	\begin{algorithmic}[1]
		\Procedure{mcubes-kernel}{$f$, $d$, $m$, $s$, $p$, $B[500 \cdot d]$, $C[500 \cdot d]$, $r$}
		\For{$m$/($b$ total threads) parallel}
		\State \textsc{Set-Random-Generator}($seed$)
		\State $I$, $E \gets 0$ \Comment{cumulative estimates of thread}

		\For{$t = 0$ to $s$}
		\State $I_t$, $ E_t \gets 0$	\Comment{estimates of sub-cube t}

		\For{$k \gets 1$ to $p$}
		\State $x[1:d] \gets$ \textsc{Generate}()
		\State $I_k, E_k \gets $ \textsc{Evaluate}($f,x$)
		\State $I_t \gets I_t + I_k$	\Comment{Accumulate sub-cube contributions}
		\State $E_t \gets E_t + E_k$
		\State $b[1:d] \gets$ \textsc{Get-Bin-ID}($x$)
		\For{$j \gets 1$ to $d$}	\Comment{Store bin contributions}
		\State \textsc{AtomicAdd}($C[b[j]]$, $I_k^2$)
		\EndFor

		\EndFor

		\State $E_t \gets$ \textsc{UpdateVariance}($E_t$, $I_t$, $p$)
		\State $I \gets I + I_t$ \Comment{update cumulative values}
		\State $E \gets E + E_t$

		\EndFor

		\State $I \gets$ \textsc{Reduce}($I$)
		\State $E \gets$ \textsc{Reduce}($E$)

		\If{thread 0 within group}
		\State \textsc{AtomicAdd}($r[0]$, $I$)
		\State \textsc{AtomicAdd}($r[1]$, $E$)
		\EndIf

		\EndFor
		\EndProcedure
	\end{algorithmic}
\end{algorithm}

The input of \textsc{mcubes-kernel}, consists of the integrand $f$ of dimensionality $d$, the number of sub-cubes $m$, sub-cube batch size $s$, number of samples per sub-cube $p$, bin bounds $B[500 \cdot d]$, bin contributions $C[500 \cdot d]$, and result $r$.
The \textsc{mcubes-kernel} outputs an estimate for the integral, variance, and the accumulated bin contributions $C$ from all samples.

All $m/b$ threads are launched in parallel at line $2$.
At line $3$ the random number generator is initialized.
Each thread processes its assigned sub-cubes serially through the loop at line $5$, and takes $p$ samples per sub-cube at line $7$.
The results from each sample are stored in $I_k$ and $E_k$ (line $9$). We accumulated those results to $I_t$ and $E_t$ (line $11$) which correspond to the sub-cube estimates. Finally, we accumulate the results from all sub-cubes in $I$ and $E$.

When evaluating each sample, we first generate a $d$-dimensional point $x$ in  line $8$. The kernel evaluates the integrand $f$ at that point(line $9$) .
Then, at line $12$, the kernel identifies the IDs of the $d$ bins used to to generate the point $x$, and stores their indices in the list $C$, at the local array $b$. Those indices are used to update the bin contribution at line $14$ through atomic addition.

The bin contribution is the square of the integral estimate $I_k$.
Then, the variance is updated at line $15$ and the error and integral estimate for the $s$ sub-cubes is accumulated in lines $16$ and $17$.

The kernel then uses block-reduction to accumulate $I$ and $E$ from each thread  at lines $18$ and $19$.
Finally, two atomic additions from thread $0$ of each thread-group leads to the computation of the total integral and error estimate from all thread-groups. These are stored in memory location $r$ at lines $21$ and $22$.

The CUDA implementation was optimized for the \emph{V100} GPU.
The kernel consisted of $128$ threads per block and utilized atomic addition for accumulating results originating from different thread blocks.
This is particularly important for updating the random-access bin contributions, which refer to $500$ bins per dimensional axis.
The reduction operations at lines $18/19$ operate on local memory and utilize warp-level primitives, though shared memory is used to accumulate the values from the different warps.

\section{Porting Process}

The maturity of the CUDA programming model along with the more widespread utilization of highly performant Nvidia GPUs make CUDA an intuitive choice for high-performance applications.
As such, PAGANI and $m$-Cubes were designed and optimized for CUDA on a V100 Nvidia GPU \cite{pagani} \cite{mcubes}.
This makes DPCT the most appropriate tool to facilitate the porting process from CUDA to DPC++.

\subsection{DPCT}

DPCT is intended to automate the majority of CUDA code migration to DPC++, instead of performing a total conversion \cite{dpct}.
In our experience as well as those reported in \cite{costanzo1}, \cite{zheming} and many others, DPCT functions exactly as intended.
An easy-to-complete conversion process requires few manual code insertions.
When manual editing is needed, DPCT displays helpful suggestions in the form of comment blocks to guide the user.
There were certain code segments that were functional but needed simple fixes to improve performance.
e.g. use of 3D \emph{nd\_item} instead of 1D equivalent.
In our experience, those cases were few and we expect that such effects will be less pronounced as oneAPI and DPCT evolve.
Expert users are anticipated to produce higher-quality implementations than automated tools, but even then DPCT greatly facilitates the porting process by automating the tedious and often error-prone translation of API calls and indexing schemes.

\subsection{Challenges}

\subsubsection{Porting Issues with Nvidia Thrust Library}
PAGANI uses Thrust to perform common parallel operations on the host side, such as reduction, dot-product, prefix sum, and finding the minimum/maximum within a list.
oneAPI provides an array of libraries to accommodate all of our required operations.
While the CUDA implementation defaults to cuRAND to generate the random samples in $m$-Cubes, we use a simple custom-made random generator in both the CUDA and oneAPI versions. This facilitates the comparison between the two implementations and avoids a comparison of random-number-generator libraries.

The usage of oneAPI libraries is more limited on the CUDA-backend.
In our experience, certain calls to library functions inserted by DPCT worked on Intel GPUs such as the P630 but were not viable on the CUDA-backend, yielding compile-time errors in certain cases such as when using \emph{mkl::stats::min\_max}.
Additionally, DPCT headers provided library calls with CUDA-like syntax in the \emph{dpct} namespace.
This worked on Intel GPUs but not on the CUDA-backend,
e.g \emph{dpct::inner\_product}, \emph{dpct::get\_device\_pointer}.
Still, we found alternatives from the oneDPL and oneMKL libraries, which allowed us to execute successfully on both the CUDA-backend and on Intel devices.


\subsubsection{Performance Degradation}
We encountered more difficulties when attempting to achieve comparable performance to the original CUDA implementation.
We observed that the NVCC and Clang compilers optimized differently when using the O3 optimization flag; we needed to set inline thresholds and to manually enable/disable loop-unrolling in certain locations.
Furthermore, code elimination in some of our benchmarking kernels that invoked mathematical functions on the device was not equivalent in the two compilers.
Additionally, we found that for our implementation, work-group reduction caused approximately 10\% slowdown in the oneAPI version compared to manual reduction using shared memory and warp-intrinsics.

Another challenge in our attempt to achieve comparable performance to CUDA was deviations in the performance of SYCL and CUDA mathematical functions.
Exponential functions displayed comparable performance on benchmark kernels.
On the contrary, we observed a slowdown of various degrees in oneAPI when using \emph{power} or trigonometric functions.
This is most likely attributed to the compilers utilizing different optimizations.
We did not use any \emph{fast-math} flags, since high accuracy is critical in numerical integration use cases.

Finally, the use of atomic addition in $m$-Cubes caused orders of magnitude slowdown on both the \textsc{mcubes-kernel} and benchmark kernels.
This was attributed to the lack of an architecture-specific flag that must be set to enable efficient atomics when supported.
After setting the Volta architecture flag, atomic addition was as performant as in the native CUDA implementation.

\subsubsection{Software Engineering Issues}
Utilization of the Catch2 testing framework and CMake was largely successful but more error-prone in oneAPI than in the case of CUDA.
Header-inclusion issues often caused non-intuitive compilation errors if headers for the oneDPL library were placed after oneMKL headers.
Setting up an environment to compile for the CUDA backend through CMake was not as easy as a build for fully-supported architectures since we require separate CMake instructions and flags when executing on Intel or Nvidia GPUs.
As the CUDA-backend for SYCL support continues, we expect that such features will be addressed.

\section{Experimental Results}

We conducted a series of experiments to evaluate the performance and correctness of the oneAPI ports relative to the optimized CUDA implementations of PAGANI and $m$-Cubes.
We used a single node with a V100 Nvidia GPU and a $2.4$ GHz Intel Xeon R Gold $6130$ CPU.
We also used the Devcloud environment to verify that the DPC++ implementations were portable and could be executed on a P630 Intel GPU.
Due to the V100 GPU having significantly more computing cores than the P630, we do not make any performance comparisons between the two GPUs.
Instead, we focus on the attainable performance of DPC++ on Nvidia hardware.

When executing the CUDA implementations, we used gcc $8.5$ and CUDA $11.6$.
For the CUDA-backend execution, we used the same environment but compiled with clang $15$, an inline threshold of $10000$, and the following compilation flags: ``-fsycl -fsycl-targets=nvptx64-nvidia-cuda -Xsycl-target-backend --cuda-gpu-arch=sm\_70''.
We verified the correctness of our ports, by comparing the results on both the Nvidia (V100) and Intel (P630) GPUs, to the results generated by the CUDA originals on a V100 GPU.

In terms of evaluating performance, we chose the same benchmark integrands originally used to evaluate PAGANI and $m$-Cubes in \cite{pagani} and \cite{mcubes}.
These functions belong to separate integrand families with features that make accurate estimation challenging.
We list those integrands in equations $1$ to $6$.
All experiments use the same integration bounds $(0,1)$ on each dimensional axis. Similar to \cite{pagani} and \cite{mcubes}, we perform multiple experiments per integrand.

We deviate from \cite{pagani} and \cite{mcubes} in that we do not execute the PAGANI and $m$-Cubes methods in their entirety.
Instead, we execute their main kernels \textsc{pagani-kernel} and \textsc{mcubes-kernel}, which is where more than $90\%$ of execution is spent.
With this approach, we can evaluate the effectiveness of each programming model in terms of offloading workloads to the device.
It allows us to separate kernel evaluation from memory management operations (allocations, copies, etc.) and library usage.
This comparison of custom kernel implementations is a better indicator of performance implications when porting CUDA codes to DPC++.

\begin{equation}
	f_{1,d}\left(x\right) = \cos\left(\sum_{i=1}^{d} i \, x_i\right)
\end{equation}

\begin{equation}
	f_{2,d}\left(x\right) = \prod_{i=1}^{d} \left(\frac{1}{50^2} + \left(x_i - 1/2\right)^2\right)^{-1}
\end{equation}

\begin{equation}
	f_{3,d}\left(x\right) = \left(1+ \sum_{i=1}^{d} i \, x_i\right)^{-d-1}
\end{equation}

\begin{equation}
	f_{4,d}\left(x\right) = \exp\left(-625 \sum_{i=1}^{d} \left(x_i-1/2\right)^2\right)
\end{equation}

\begin{equation}
	f_{5,d}\left(x\right) =	\exp\left(-10  \sum_{i=1}^{d} | x_i - 1/2 |\right)
\end{equation}

\begin{equation}
	f_{6,d}\left(x\right) =
	\begin{cases}
		\exp\left(\sum_{i=1}^d \left(i+4\right) x_i\right) & \text{ if } x_i < \left(3+i\right)/10 \\
		0                                                  & \text{otherwise}
	\end{cases}
\end{equation}

\subsection{Offloading Mathematical Computations to Kernels}

A critical stage in \textsc{pagani-kernel} and \textsc{m-Cubes-kernel} is the invocation of the integrand at various $d$-dimensional points.
Integrands with trigonometric or exponential functions and table look-ups will have larger execution times compared to other simple integrands that only contain basic mathematical operations.
To attain satisfactory performance, both the invocation of the integrand functions and the remaining operations within the kernels must achieve comparable performance to the CUDA implementation.

We tested the efficiency of the integrand oneAPI implementations with a simple kernel that performs a series of invocations on many $d$-dimensional points.
The points are randomly generated on the host and then copied to device memory.
Each thread invokes the integrand serially $1$ million times and writes its accumulated results to global memory.
Writing the results prevents the NVCC and Clang compilers from disregarding the integrand computations due to optimization.

We first tested simple integrands that contained only a particular function such as \emph{sin, pow, powf, sycl::exp, sycl::pow, sycl::pown}.
We invoked these mathematical functions with $d$ arguments that comprise each $d$-dimensional point.
We did not use fast-math flags as accuracy is critical in numerical integration.
We observed small but consistent penalties of at most $2\%$ when invoking the \emph{power} and \emph{exponential} functions.
On the contrary, trigonometric functions are approximately $40\%$ slower on the CUDA backend.

We performed the same experiment on the six benchmark integrands for dimensions $5$ to $8$.
We summarize the results in Table \ref{table:benchark-integrands-invocation}.
The timings in CUDA and oneAPI columns are the means of 10 kernel executions per integrand.
The ratio of those timings shows that the oneAPI version is at most $4\%$ slower.
The largest penalty is observed in the $f1$ integrand which makes use of the $cos$ function.
The remaining integrands only make use of \emph{exponential} and \emph{power} functions and yield small penalties.

These experiments on the execution time of the integrand invocations demonstrate that the user-defined computations do not display significant performance penalties.
The one exception is the extended use of trigonometric functions.
None of the benchmark integrands make extended use of trigonometric functions ($f1$ has one call to $cos$ per invocation).
As such, we do not expect any slowdown larger than $5\%$ in either PAGANI or $m$-Cubes to be attributed to the integrand implementations.

\begin{table}[]
	\centering
	\caption{mean ($\mu$) and standard deviation ($\sigma$) of execution times for invoking $5-8D$ benchmark integrands}
	\label{table:benchark-integrands-invocation}
	\renewcommand{\arraystretch}{1.5}
	\begin{tabular}{p{1.5cm}p{2.5cm}p{2.5cm}p{2cm}p{2cm}p{1cm}}

		\hline
		id & $\mu$ CUDA (ms) & $\mu$ oneAPI (ms) & $\sigma$ CUDA & $\sigma$ oneAPI & $\frac{\mu \; oneAPI}{\mu \;CUDA}$ \\
		\hline
		f1 & 1866.4    & 1952.4      & 13.3      & 21.4        & 1.04                  \\
		f2 & 8413.9    & 8487.3      & 5012.5    & 5042.9      & 1.009                 \\
		f3 & 1812.4    & 1828.3      & 18.5      & 27.1        & 1.009                 \\
		f4 & 11416.1   & 11410.1     & 2184.9    & 2148.1      & 0.99                  \\
		f5 & 634.3     & 654.4       & 73.5      & 67.3        & 1.03                  \\
		f6 & 300.4     & 300.8       & 32.05     & 32.6        & 1.001                 \\  \hline
	\end{tabular}
\end{table}

\subsection{Benchmark Integrands Performance Comparison}

Another set of experiments involved the invocation of the \textsc{pagani-kernel} and \textsc{mcubes-kernel} on the benchmark integrands.
To address different degrees of computational intensity, we vary the number of thread-blocks used to launch the kernels.
For the \textsc{mcubes-kernel}, we achieve this effect by varying the required number of samples per iteration in the range ($1e8, 3e9$).
This leads to different block sizes per kernel.
For \textsc{pagani-kernel}, the number of thread blocks corresponds to the number of regions being processed.
We perform high-resolution uniform splits to generate region lists of different sizes and supply them to the \textsc{pagani-kernel} for evaluation.

We report the penalty of using oneAPI for the benchmark integrands, in the ratio columns of Tables \ref{table:mcubes8D} and \ref{table:pagani8D}.
We used four thread-block sizes for each integrand for the kernel executions.
Each kernel configuration (number of thread groups) was repeated $100$ times to provide a statistical mean and standard deviation for the execution times.

Across our experiments, the average execution time ratio ($\frac{oneAPI}{CUDA}$) is in the range ($0-10\%$).
The $f2$ and $f4$ integrands which make repeated use of the \emph{power} function display the largest performance penalties for both PAGANI and $m$-Cubes.
It is worth noting that both $f2$ and $f4$ display the largest execution times among the benchmark integrands for both integrators.

\begin{table}[ht]
	\centering
	\caption{$m$-Cubes: mean ($\mu$) and standard deviation ($\sigma$) of execution times for 8D benchmark integrands}
	\label{table:mcubes8D}
	\renewcommand{\arraystretch}{1.5}
	\begin{tabular}{p{1.5cm}p{2.5cm}p{2.5cm}p{2cm}p{2cm}p{1cm}}
		\hline
	id & $\mu$ CUDA (ms) & $\mu$ oneAPI (ms) & $\sigma$ CUDA & $\sigma$ oneAPI & $\frac{\mu \; oneAPI}{\mu \;CUDA}$ \\
		\hline
		f1 & 286.7     & 286.7       & 2.1       & 0.9         & 1.0                   \\
		f2 & 402.1     & 443.1       & 2.6       & 0.9         & 1.1                   \\
		f3 & 284.5     & 285.8       & 1.6       & 1.4         & 1.0                   \\
		f4 & 385.7     & 423.5       & 2.4       & 0.5         & 1.1                   \\
		f5 & 284.3     & 285.9       & 2.1       & 1.7         & 1.0                   \\
		f6 & 283.8     & 285.4       & 1.9       & 1.6         & 1.0                   \\  \hline
	\end{tabular}
\end{table}

\begin{table}[ht]
	\centering
	\caption{PAGANI: mean ($\mu$) and standard deviation ($\sigma$) of execution times for 8D benchmark integrands}
	\label{table:pagani8D}
	\renewcommand{\arraystretch}{1.5}
		\begin{tabular}{p{1.5cm}p{2.5cm}p{2.5cm}p{2cm}p{2cm}p{1cm}}
		\hline
		id & $\mu$ CUDA (ms) & $\mu$ oneAPI (ms) & $\sigma$ CUDA & $\sigma$ oneAPI & $\frac{\mu \; oneAPI}{\mu \;CUDA}$ \\
		\hline
		f1 & 172.3     & 177.5       & 0.9       & 1.2         & 1.02                  \\
		f2 & 1500.4    & 1651.0      & 0.3       & 2.1         & 1.1                   \\
		f3 & 286.4     & 290.7       & 0.8       & 0.4         & 1.01                  \\
		f4 & 1434.7    & 1524.9      & 0.4       & 1.9         & 1.06                  \\
		f5 & 166.5     & 170.7       & 0.6       & 0.4         & 1.03                  \\
		f6 & 136.8     & 139.4       & 0.4       & 0.2         & 1.02                  \\
		\hline
	\end{tabular}
\end{table}

\subsection{Simple Integrands Performance Comparison}


In addition to the benchmark integrands, we also evaluate integrands that only perform a summation of the arguments ($\sum_{i=1}^{d} x_{i}$) where $d$ is the number of dimensions.
This avoids any bias in the comparison by avoiding mathematical functions that could either call different implementations, cause differences in register usage or lead to different optimizations.
The ratios in Tables \ref{table:mcubes-addition-integrands} and \ref{table:pagani-addition-integrands}, display timings on addition integrands for dimensions five to eight.
Once more, we observe penalties smaller than $10\%$ and for both integrators these penalties decrease on higher dimensionalities.



\begin{table}[ht]
	\centering
	\caption{$m$-Cubes: mean ($\mu$) and standard deviation ($\sigma$) of execution times for addition integrands ($\sum_{i=1}^{d} x_{i}$)}
	\label{table:mcubes-addition-integrands}
	\renewcommand{\arraystretch}{1.5}
	\begin{tabular}{p{1.5cm}p{2.5cm}p{2.5cm}p{2cm}p{2cm}p{1cm}}\hline
		id & $\mu$ CUDA (ms) & $\mu$ oneAPI (ms) & $\sigma$ CUDA & $\sigma$ oneAPI & $\frac{\mu \; oneAPI}{\mu \;CUDA}$ 
		\\ \hline
		5D & 206.1     & 214.5       & 2.1       & 1.7         & 1.04                  \\
		6D & 214.1     & 217.2       & 2.2       & 1.0         & 1.01                  \\
		7D & 234.1     & 235.2       & 1.8       & 0.9         & 1.005                 \\
		8D & 284.7     & 285.7       & 1.9       & 1.9         & 1.005                 \\
		\hline
	\end{tabular}
\end{table}

\begin{table}[ht]
	\centering
	\caption{PAGANI: mean ($\mu$) and standard deviation ($\sigma$) of execution times for addition integrands ($\sum_{i=1}^{d} x_{i}$)}
	\label{table:pagani-addition-integrands}
	\renewcommand{\arraystretch}{1.5}
	\begin{tabular}{p{1.5cm}p{2.5cm}p{2.5cm}p{2cm}p{2cm}p{1cm}}\hline
		id & CUDA (ms) & oneAPI (ms) & Std. CUDA & Std. oneAPI & $\frac{oneAPI}{CUDA}$ \\ \hline
		5D & 1.5       & 1.7         & 0.05      & 0.06        & 1.1                   \\
		6D & 24.8      & 26.7        & 0.3       & 1.4         & 1.1                   \\
		7D & 129.8     & 131.6       & 0.7       & 0.2         & 1.01                  \\
		8D & 137.4     & 137.6       & 1.3       & 1.0      & 1.001                 \\
		\hline
	\end{tabular}
\end{table}

\subsection{Factors Limiting Performance}

Both \textsc{pagani-kernel} and \textsc{mcubes-kernel}, are compute bound, performing thousands of computations for each byte of accessed memory.
For compute-bound kernels, the number of registers per thread is a factor limiting the number of concurrent threads that can be executed; the amount of shared memory and registers per thread limit warp/work-group occupancy, which in turn degrades performance. 

\begin{figure}
	\centering
	\begin{minipage}{0.45\textwidth}
		\centering
		\includegraphics[width=0.9\textwidth]{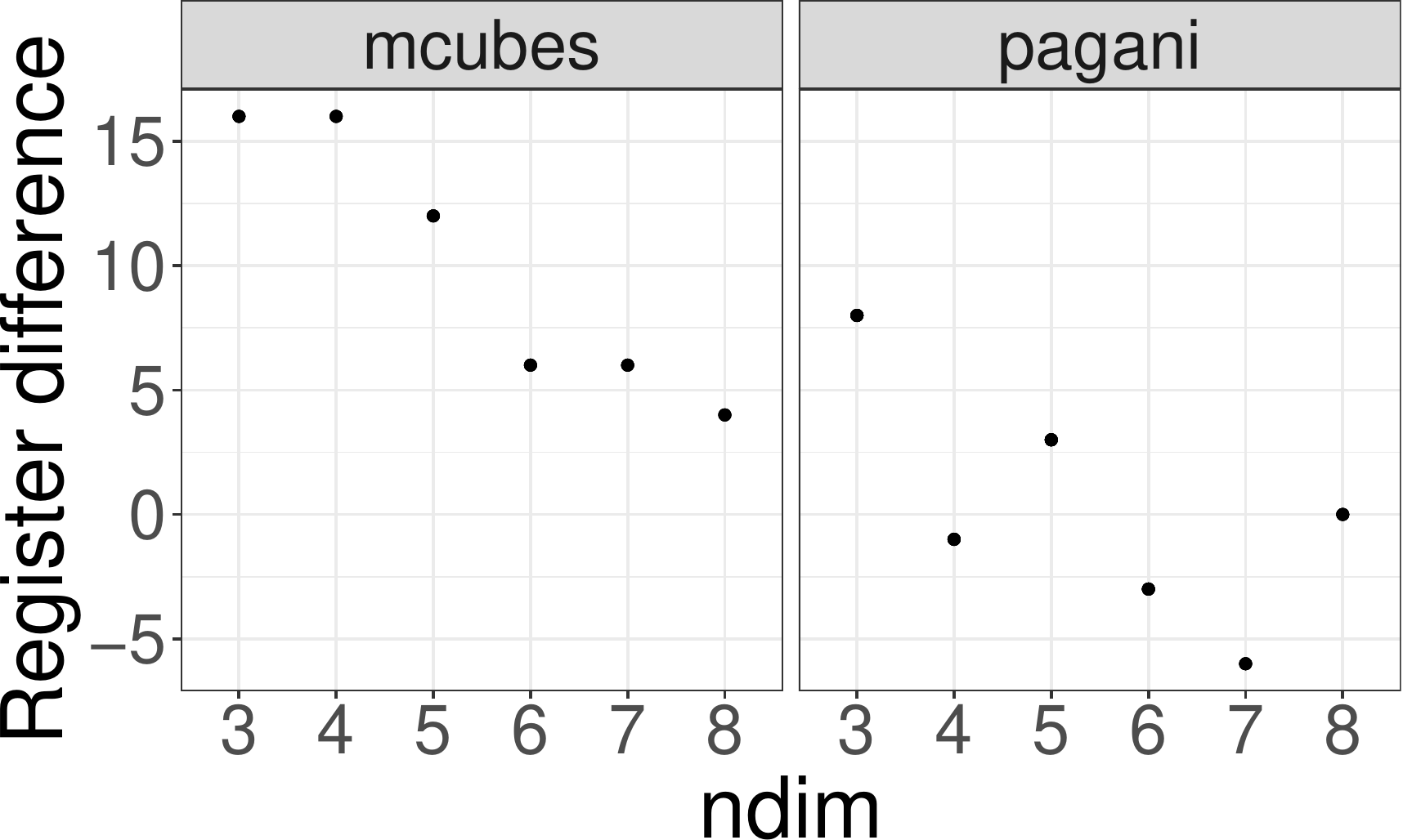} 
		\caption{Register difference on simple addition integrands ($\sum_{i=1}^{d} x_{i}$). The y-axis displays the number of additional registers per thread in the DPC++ implementation.}
		\label{fig:addition_registers}

	\end{minipage}\hfill
	\begin{minipage}{0.45\textwidth}
		\centering
		\includegraphics[width=0.9\textwidth]{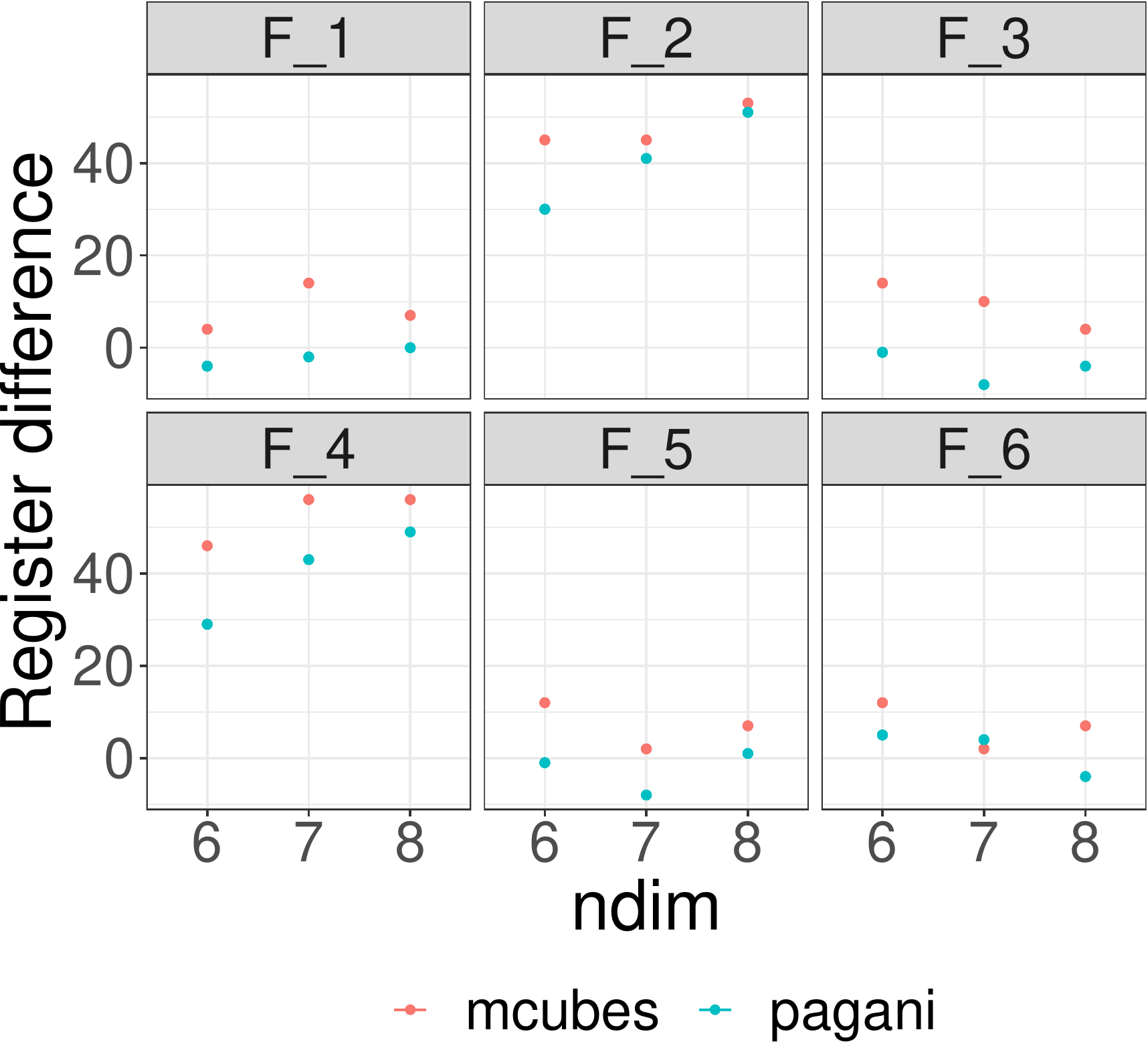} 
		\caption{Register difference on the benchmark integrands. The y-axis displays the number of additional registers per thread in the DPC++ implementation.}
		\label{fig:F_registers}

	\end{minipage}
\end{figure}

In most cases, the oneAPI implementations assigned more registers to each thread compared to their CUDA equivalents.
We illustrate the magnitude of this difference in registers per thread in Figures \ref{fig:addition_registers} and \ref{fig:F_registers}.
We observe the largest difference in integrands $f2$ and $f4$, which make extended use of the \emph{power function}.
It is the same functions that display the two largest execution time penalties for the benchmark integrands in Tables \ref{table:mcubes8D} and \ref{table:pagani8D}.

We observe a similar pattern on the simple addition integrands (Table \ref{table:mcubes-addition-integrands} and \ref{table:pagani-addition-integrands}). 
In those cases, there are no mathematical functions (\emph{pow}, \emph{exp}, etc.) and the integrands only perform a summation. 
The difference in registers decreases on higher dimensions, leading to degraded performance on low dimensions. 
This is evident in tables \ref{table:mcubes-addition-integrands} and \ref{table:pagani-addition-integrands} where higher-dimensional integrands have smaller values in the $\frac{oneAPI}{CUDA}$ column.
The same pattern is observed for the benchmark integrands, where the high dimensional versions perform better than the low dimension equivalents. It can be seen in Figure \ref{fig:addition_registers}, that this effect is more prominent in $m$-Cubes, since it displays a larger deviation across all dimensions.
These observations lead us to believe that register difference and its effect on occupancy is the main reason behind the performance degradation.

\section{Conclusion}

We presented our experience of porting two numerical integration implementations, PAGANI and $m$-Cubes, from CUDA to DPC++ .
We utilized Intel's DPCT to automate the conversion process from CUDA to DPC++ and successfully attained the capability to execute the same implementation on both Intel and Nvidia GPUs.
We experimented with various workloads consisting of different mathematical functions.
We found that the assigned registers per thread can deviate in oneAPI and CUDA codes. This affects occupancy which in turn can negatively impact performance, particularly in compute-bound kernels.
We faced additional challenges with mapping library calls to oneAPI equivalents, matching compiler optimizations of NVCC with Clang, and using build and testing libraries like CMake and Catch2.
We addressed those challenges and demonstrated that the performance penalty of using oneAPI ports instead of optimized CUDA implementations can be limited to $10\%$ on Nvidia GPUs. Additionally, numerous cases exhibited comparable performance to the original CUDA implementations, with execution time differences in the $1-2\%$ range. 
We compared oneAPI and CUDA implementations on the same Nvidia V100 GPU. We were able to execute on an Intel P630 GPU but we did not compare these timings with those on the V100 GPU due their significant difference in computing power. In the future, we plan to execute on the high end Intel Ponte Vecchio GPU and compare performance metrics with Nvidia high end GPUs such as A100. 
 
The vast array of libraries, ease of portability, and small margin of performance degradation, make oneAPI an appropriate software solution for the use case of numerical integration.

\section{Acknowledgements}

The authors would like to thank Intel Corporation and Codeplay for providing technical support in the conversion process. The authors are also grateful for the support of the Intel oneAPI Academic Center of Excellence at Old Dominion University.

Work supported by the Fermi National Accelerator Laboratory, managed and operated
by Fermi Research Alliance, LLC under Contract No. DE-AC02-07CH11359 with the U.S.
Department of Energy. The U.S. Government retains and the publisher, by accepting the article for publication, acknowledges that the U.S. Government retains a non-exclusive, paid-up, irrevocable, world-wide license to publish or reproduce the published form of this manuscript, or allow others to do so, for U.S. Government purposes. FERMILAB-CONF-23-007-LDRD-SCD.

We acknowledge the support of Jefferson Lab grant to Old Dominion University 16-347. Authored by Jefferson Science Associates, LLC under U.S. DOE Contract No. DE-AC05-06OR23177 and DE-AC02- 06CH11357.

\bibliography{refs}

\begin{thebibliography}{10}
\providecommand{\url}[1]{\texttt{#1}}
\providecommand{\urlprefix}{URL }
\providecommand{\doi}[1]{https://doi.org/#1}

\bibitem{openmp}
\url{https://www.openmp.org/wp-content/uploads/OpenMP-API-Specification-5.0.pdf}

\bibitem{alcf}
Argonne leadership computing facility,
  \url{https://www.alcf.anl.gov/support-center/aurora/sycl-and-dpc-aurora#:~:text=DPC%2B%2B%20(Data%20Parallel%20C,versions%20of%20the%20SYCL%20language}

\bibitem{computecpp}
Computecpp™ community edition,
  \url{https://developer.codeplay.com/products/computecpp/ce/2.11.0/guides/#computecpp}

\bibitem{dpct}
Migrate cuda* to dpc++ code: Intel® dpc++ compatibility tool,
  \url{https://www.intel.com/content/www/us/en/developer/tools/oneapi/dpc-compatibility-tool.html#gs.lx007q}

\bibitem{openacc}
What is openacc?, \url{https://www.openacc.org/}

\bibitem{des}
et~al., G.: Dark energy survey year 3 results: Redshift calibration of the
  maglim lens sample from the combination of sompz and clustering and its
  impact on cosmology  (2022)

\bibitem{twophase}
Arumugam, K., Godunov, A., Ranjan, D., Terzic, B., Zubair, M.: A memory
  efficient algorithm for adaptive multidimensional integration with multiple
  gpus. In: 20th Annual International Conference on High Performance Computing.
  pp. 169--175. IEEE (2013)

\bibitem{dpcpp}
Ashbaugh, B., Bader, A., Brodman, J., Hammond, J., Kinsner, M., Pennycook, J.,
  Schulz, R., Sewall, J.: Data parallel c++: Enhancing sycl through extensions
  for productivity and performance. In: Proceedings of the International
  Workshop on OpenCL. IWOCL '20, Association for Computing Machinery, New York,
  NY, USA (2020). \doi{10.1145/3388333.3388653},
  \url{https://doi.org/10.1145/3388333.3388653}

\bibitem{cosmosis}
Bridle, S., Dodelson, S., Jennings, E., Kowalkowski, J., Manzotti, A., Paterno,
  M., Rudd, D., Sehrish, S., Zuntz, J.: Cosmosis: a system for mc parameter
  estimation. Journal of Physics: Conference Series  \textbf{664}(7),  072036
  (dec 2015). \doi{10.1088/1742-6596/664/7/072036},
  \url{https://dx.doi.org/10.1088/1742-6596/664/7/072036}

\bibitem{kokkos}
Carter~Edwards, H., Trott, C.R., Sunderland, D.: Kokkos: Enabling manycore
  performance portability through polymorphic memory access patterns. Journal
  of parallel and distributed computing  \textbf{74}(12),  3202--3216 (2014)

\bibitem{CASTANO2022120}
Castaño, G., Faqir-Rhazoui, Y., García, C., Prieto-Matías, M.: Evaluation of
  intel's dpc++ compatibility tool in heterogeneous computing. Journal of
  Parallel and Distributed Computing  \textbf{165},  120--129 (2022).
  \doi{https://doi.org/10.1016/j.jpdc.2022.03.017},
  \url{https://www.sciencedirect.com/science/article/pii/S0743731522000727}

\bibitem{stencil_code}
Christgau, S., Steinke, T.: Porting a legacy cuda stencil code to oneapi. In:
  2020 IEEE International Parallel and Distributed Processing Symposium
  Workshops (IPDPSW). pp. 359--367 (2020). \doi{10.1109/IPDPSW50202.2020.00070}

\bibitem{costanzo3}
Costanzo, M., Rucci, E., García-Sánchez, C., Naiouf, M., Prieto-Matías, M.:
  Migrating cuda to oneapi: A smith-waterman case study. In: Bioinformatics
  and Biomedical Engineering, pp. 103--116. Lecture Notes in Computer Science,
  Springer International Publishing, Cham (2022)

\bibitem{costanzo1}
Costanzo, M., Rucci, E., Sanchez, C.G., Naiouf, M.: Early experiences migrating
  cuda codes to oneapi  (2021)

\bibitem{costanzo2}
Costanzo, M., Rucci, E., Sánchez, C.G., Naiouf, M., Prieto-Matías, M.:
  Assessing opportunities of sycl and intel oneapi for biological sequence
  alignment  (2022)

\bibitem{sycl}
Doerfert, J., Jasper, M., Huber, J., Abdelaal, K., Georgakoudis, G., Scogland,
  T., Parasyris, K.: Breaking the vendor lock-performance portable programming
  through openmp as target independent runtime layer. Tech. rep., Lawrence
  Livermore National Lab.(LLNL), Livermore, CA (United States) (2022)

\bibitem{bioinf_kernel}
Haseeb, M., Ding, N., Deslippe, J., Awan, M.: Evaluating performance and
  portability of a core bioinformatics kernel on multiple vendor gpus. In: 2021
  International Workshop on Performance, Portability and Productivity in HPC
  (P3HPC). pp. 68--78 (2021). \doi{10.1109/P3HPC54578.2021.00010}

\bibitem{zheming}
Jin, Z., Vetter, J.: Evaluating cuda portability with hipcl and dpct. In: 2021
  IEEE International Parallel and Distributed Processing Symposium Workshops
  (IPDPSW). pp. 371--376 (2021). \doi{10.1109/IPDPSW52791.2021.00065}

\bibitem{kanzaki}
Kanzaki, J.: Monte carlo integration on gpu. The European physical journal. C,
  Particles and fields  \textbf{71}(2), ~1--7 (2011)

\bibitem{PETERLEPAGE1978192}
{Peter Lepage}, G.: A new algorithm for adaptive multidimensional integration.
  Journal of Computational Physics  \textbf{27}(2),  192--203 (1978).
  \doi{https://doi.org/10.1016/0021-9991(78)90004-9},
  \url{https://www.sciencedirect.com/science/article/pii/0021999178900049}

\bibitem{physics_app}
Ranjan, N., Terzić, B., Krafft, G., Petrillo, V., Drebot, I., Serafini, L.:
  Simulation of inverse compton scattering and its implications on the
  scattered linewidth. Physical review. Accelerators and beams  \textbf{21}(3),
   030701 (2018)

\bibitem{pagani}
Sakiotis, I., Arumugam, K., Paterno, M., Ranjan, D., Terzi\'{c}, B., Zubair,
  M.: PAGANI: A Parallel Adaptive GPU Algorithm for Numerical Integration.
  Association for Computing Machinery, New York, NY, USA (2021),
  \url{https://doi.org/10.1145/3458817.3476198}

\bibitem{mcubes}
Sakiotis, I., Arumugam, K., Paterno, M., Ranjan, D., Terzić, B., Zubair, M.:
  m-cubes: An efficient and portable implementation of multi-dimensional
  integration for gpus. In: High Performance Computing, pp. 192--209. Lecture
  Notes in Computer Science, Springer International Publishing, Cham (2022)

\bibitem{opencl}
Stone, J.E., Gohara, D., Shi, G.: Opencl: A parallel programming standard for
  heterogeneous computing systems. Computing in science \& engineering
  \textbf{12}(3),  66--73 (2010)

\bibitem{opencl_cuda}
Su, C.L., Chen, P.Y., Lan, C.C., Huang, L.S., Wu, K.H.: Overview and comparison
  of opencl and cuda technology for gpgpu. In: 2012 IEEE Asia Pacific
  Conference on Circuits and Systems. pp. 448--451 (2012).
  \doi{10.1109/APCCAS.2012.6419068}

\bibitem{linalg}
Tsai, Y.M., Cojean, T., Anzt, H.: Porting sparse linear algebra to intel gpus.
  In: Euro-Par 2021: Parallel Processing Workshops, pp. 57--68. Lecture Notes
  in Computer Science, Springer International Publishing, Cham (2022)

\bibitem{boris_p_pusher}
Volokitin, V., Bashinov, A., Efimenko, E., Gonoskov, A., Meyerov, I.: High
  performance implementation of boris particle pusher on dpc++. a first look at
  oneapi. In: Lecture Notes in Computer Science (including subseries Lecture
  Notes in Artificial Intelligence and Lecture Notes in Bioinformatics), pp.
  288--300. Lecture Notes in Computer Science, Springer International
  Publishing, Cham (2021)

\bibitem{sycl_overview}
Wong, M., Liber, N., Bassini, S., Richards, A., Butler, M., McVeigh, J., Cook,
  B., Sugimoto, H., Cordoba, C., Fahringer, T., et~al.: Sycl - c++
  single-source heterogeneous programming for acceleration offload (Jan 2014),
  \url{https://www.khronos.org/sycl/}

\bibitem{zmc}
Wu, H.Z., Zhang, J.J., Pang, L.G., Wang, Q.: Zmcintegral: A package for
  multi-dimensional monte carlo integration on multi-gpus. Computer Physics
  Communications  \textbf{248},  106962 (2020).
  \doi{https://doi.org/10.1016/j.cpc.2019.106962},
  \url{https://www.sciencedirect.com/science/article/pii/S0010465519303121}

\bibitem{alpaka}
Zenker, E., Worpitz, B., Widera, R., Huebl, A., Juckeland, G., Knüpfer, A.,
  Nagel, W.E., Bussmann, M.: Alpaka - an abstraction library for parallel
  kernel acceleration. In: arXiv.org. Cornell University Library, arXiv.org,
  Ithaca (2016)

\end{thebibliography}
\end{document}